\documentclass{PoS}

\usepackage{booktabs}
\usepackage{amsmath}
\usepackage{amsfonts}
\usepackage{amssymb}
\usepackage{mathrsfs}
\usepackage{latexsym}
\usepackage{dsfont}
\usepackage{bm}
\usepackage{revsymb}
\usepackage{subfig}
\input{ogbabarsym}

\title{Study of baryonic decays of B mesons at \babar}

\ShortTitle{Study of baryonic decays of B mesons at Babar}

\author{\speaker{Oliver Gruenberg} (for the \babar ~collaboration)\\
        Rostock University\\
        E-mail: \email{oliver.gruenberg@uni-rostock.de}}

\abstract{ \vspace{3em}
           \begin{center}\includegraphics[width=0.4\textwidth]{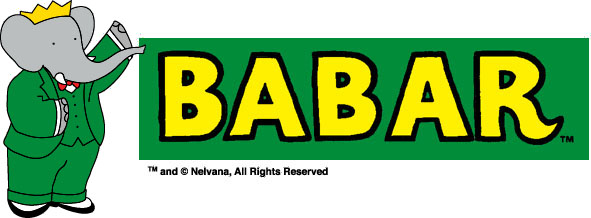}\end{center}
           \vspace{2em}
           We report on recent \babar ~analyses of the baryonic \B decays \thdec, \mydec, \mydecII and \cvdec.
           The used data sample contains $471\times10^{6}$ \BB pairs that were generated in the process 
           \epem\to\upsbb and collected with the \babar ~detector at the \pep2 ~storage ring at SLAC.
           We find $\thbr = (12.3\pm0.5\ustat\pm0.7\usyst\pm3.2_{\rm\Lambda_c})\times10^{-3}$, 
           $\mybr = (2.98\pm0.16\ustat\pm0.15\usyst\pm0.77_{\rm\Lambda_c})\times10^{-3}$, where the last
           uncertainty is due to $\BR(\Lcp\to\proton\Km\pip)$, respectively. For the decay $\mydecII$ 
           we see no events and set an upper limit 
           $\mybrII\times\frac{\BR(\Lcp\to\proton\Km\pip)}{0.050} <2.8\times10^{-6}$ at $90\,\%$ CL, where we have 
           normalized $\BR(\Lcp\to\proton\Km\pip)$ to the world average value. There is evidence for the decay 
           $\cvdec$ and we measure $\cvbr = (9.8^{+2.9}_{-2.6}\,\ustat\pm1.9\usyst)\times10^{-6}$ corresponding 
           to a significance of $3.4\sigma$.
          }

\FullConference{XV International Conference on Hadron Spectroscopy \\
		 4-8/11/2013\\
		 Nara, Japan}

\begin{document}

\vspace*{-3.5em}

\section{Introduction}

\noindent
From ARGUS measurements published in 1992 it is known that approximately $7\%$ of all \B decays 
have baryons among their final states \cite{argus}. This is a substantial fraction which has motivated 
a large number of studies on exclusive decay modes that were published in the last twenty years, 
mainly by \babar, Belle, and CLEO. However, until today the branching fractions of these modes sum up 
to only $(0.53\pm0.06)\%$ for \Bz and $(0.85\pm0.15)\%$ for \Bp \cite{RPP}, and leaving us the 
until now unsolved $\B\to baryon$ puzzle.

Due to its large mass, \B mesons are able to decay to a large spectrum of baryons with different 
flavours such as $\proton, \Lambda, \Lcp, \Xi_{cs}$, as well as their isospin partners and 
excited states. Hence exclusive modes of baryonic \B decays are a good place to search for exotic 
baryons. On the other hand baryonic \B decays allow better understanding of hadronization into baryons 
at low $\qsq$ values. One feature that has been observed in a number of decay channels is an enhanced baryon 
production at the baryon-antibaryon mass threshold and the dependence of the branching fraction on the multiplicity; 
it can be stated that $\BR(\B\to N\Nbar)<\BR(\B\to N\Nbar\pi)<\BR(\B\to N\Nbar\pi\pi)$. 
In addition, it has been found that baryonic \B decays may have complex resonant substructures and that the branching fraction of 
decays of the sort $\B\to\Lcp\antiproton + n\cdot \pi$ with $n=1,2,3$ have large contributions from intermediate states 
including baryon resonances.

In the following we present recent \babar ~results on studies of the baryonic \B decays
\footnote{The use of charge conjugate decays is implied throughout this text.}
\thdec, \mydecII, \mydec, ~and \cvdec using the full \babar ~dataset with an integrated 
luminosity of $426\invfb$, which corresponds to $471\times10^{6}$ \BB pairs.

\section{\thdec~\cite{thomas}}

\vspace*{-1em}

\begin{figure}[ht!]
\begin{minipage}{0.49\linewidth}
\noindent
In the analysis the total signal yield is extracted in the plane of the reconstructed \B mass $m_{\rm inv}$
and $m(\Lcp\pi)$ (Fig.~\ref{fig:B2Lc2pi:mB2}) to seperate contributions that include intermediate states with 
$\Sigma_{\rm c}(2455,2520)^{0,++}\to\Lcp\pi^{-,+}$ (four exclusive modes) from the remaining non-$\Sigma_{\rm c}$ signal.
It shows that decay modes with a $\Scpp$ are prefered w.r.t.~$\Scz$. Summing up, the four resonant three-body modes amount 
to about one third of the total branching fraction. The measured branching fraction is
$\thbr = (12.3\pm0.5\ustat\pm0.7\usyst\pm3.2_{\rm\Lambda_c})\times10^{-3} $

\end{minipage}
\rule{0.5em}{0em}
\begin{minipage}{0.49\linewidth}
\begin{center}
\includegraphics[width=0.99\textwidth]{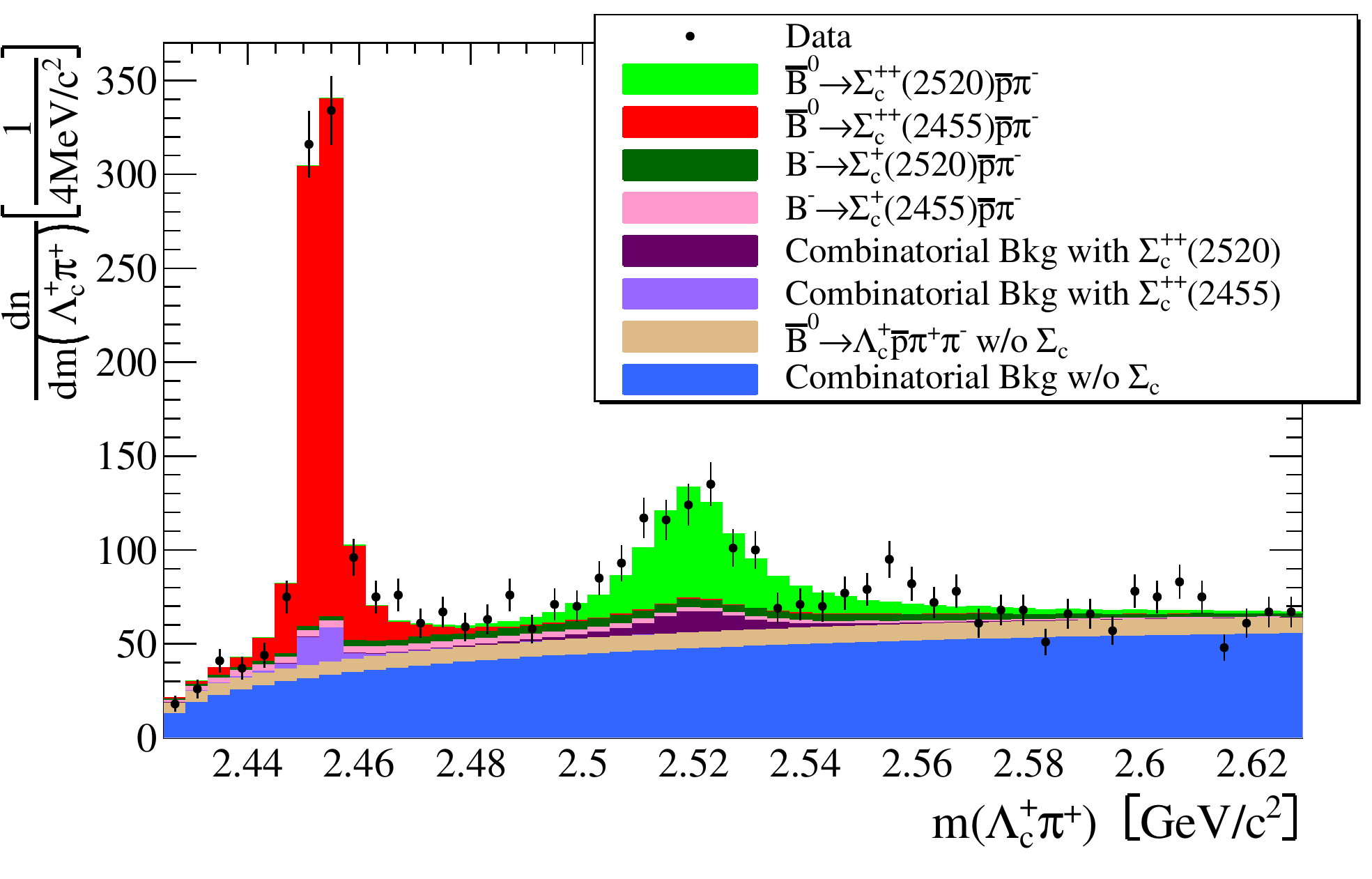}
\caption{The $m(\Lcp\pip)$ distribution in data (points with errors) and the different signal and background contributions fitted to the data.}
\label{fig:B2Lc2pi:mB2}
\end{center}
\end{minipage}
\end{figure}

\noindent
Beside the $\Sigma_{\rm c}(2455)$ and $\Sigma_{\rm c}(2520)$ resonances that are accounted for in the analysis, other intermediate states 
are visible in the invariant mass spectra of $\Lcp\pi$, $\antiproton\pi$ and $\pip\pim$. There are suggestions 
for the $\Sigma_{\rm c}(2800)$, $\Deltabar^{--}$ and $\rho(770)$ \cite{thomas}. Similar to the situation 
of $\Sigma_{\rm c}(2455)$ and $\Sigma_{\rm c}(2520)$, the signal for the $\Sigma_{\rm c}(2800)^{++}$ is more pronounced 
than for the $\Sigma_{\rm c}(2800)^{0}$.

\clearpage

\section{\mydecII~\cite{oliver2}}

\vspace*{0.5em}

\begin{figure}[ht!]
\begin{minipage}{0.48\linewidth}
\vspace{-2em}
This decay is interesting in comparison to \thdec because \mydecII cannot have intermediate states, 
for example due to $\Sigma_{\rm c}$ or $\Deltabar$ resonances, and has a very small phase space

$$ \frac{\int d{\rm PS}(\mydecII)}{\int d{\rm PS}(\thdec)}\approx \frac{1}{1400} \,. $$ 

\noindent
The ratio of \mybrII over \thbr may be larger than the phase space factor of $1/1400$ 
due to the the fact that the signal of \mydecII is limited to a small phase space 
and the invariant masses of $\Lcp\antiproton$ and $\proton\antiproton$ are generally low and
hence the hadronization into baryons may be ruled by threshold enhancement effects.

In our analysis we define a signal region in the plane of the reconstructed \B mass \mB and 
the energy-substituted mass $\mes\approx\sqrt{(\sqrt{s}/2)^2 - p^2_{\rm B}}$, with $\sqrt{s}$ the center-of-mass
energy (the exact formula takes into account the boost of the initial \epem system in the laboratory frame). 
The signal region is an ellipse that uses the mean and 
the standard deviation of \mes and \mB extracted from fits to the signal MC sample 
(Fig.~\ref{fig:B2Lc3p:mBmES}).
\end{minipage}
\rule{0.02\linewidth}{0mm}
\begin{minipage}{0.48\linewidth}
\vspace*{-1.5em}
\includegraphics[width=0.99\textwidth]{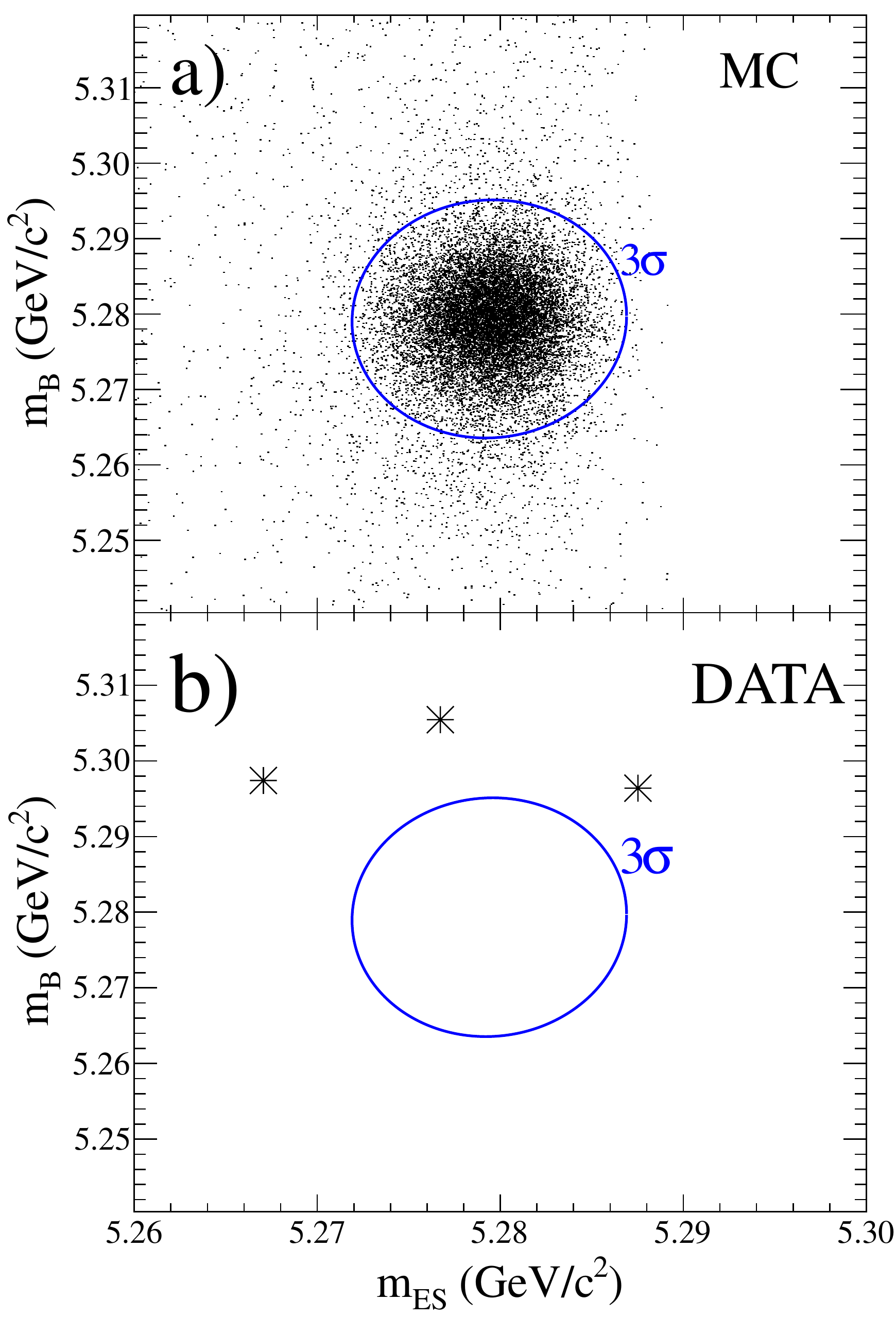}
\caption{The \mes vs. \mB distribution of selected events in (a) signal MC and (b) data.
No signal candidates are observed within the signal region of the data sample.}
\label{fig:B2Lc3p:mBmES}
\end{minipage}
\end{figure}

\noindent
We find no events in the signal region in the \babar ~data and set an upper limit on the branching fraction 

\vspace{-0.5em}
$$\mybrII\times\frac{\BR(\Lcp\to\proton\Km\pip)}{0.050} <2.8\times10^{-6}~\text{at}~90\,\%~\text{CL},$$

\noindent
where we have normalized $\BR(\Lcp\to\proton\Km\pip)$ to the world average value \cite{RPP}.
We compare this upper limit to $\thbr_{\rm non-res}\approx 50\%\cdot\thbr = 6\cdot10^{-4}$,
which is $\thbr$ without contributions from intermediate states with $\Sigma_{\rm c}(2455,2520)^{0,++}\to\Lcp\pi^{-,+}$ 
and estimated contributions from intermediate states with $\Deltabar, N^*$ and $\rho$ resonances that are visible in the 
spectra of $m(\antiproton\pipm)$ and $m(\pip\pim)$. We find that there is no significantly enhanced hadronization 
into baryons due to threshold enhancement effects in the small phase space of the decay. 
This can be quantified by an effective enhancement factor that is smaller than seven.

$$ \frac{\mybrII}{\thbr_{\rm non-res}}\lesssim 7\times\frac{1}{1400}$$ 

\clearpage

\section{\mydec~\cite{oliver1}}

\vspace*{-1em}

\begin{figure}[ht!]
\begin{minipage}{0.48\linewidth}
This decay is a resonant subchannel of \mydecF, which has the largest known branching fraction among all baryonic \B decays, 
and its studies may allow better understanding of the resonant substructure of baryon production in \B decays. 
The \Scpp ~is reconstructed as $\Scpp\to(\proton\Km\pip)_{\Lambda_{\rm c}}\pip$.
In the \babar ~data we find $787\pm43$ signal events with a reconstruction efficiency of $11.3\%$
determined from signal MC. Therefore we derive $\mybr = (2.98\pm0.16\ustat\pm0.15\usyst\pm0.77_{\Lambda_{\rm c}})\cdot10^{-4}$ 
which is about $14\%$ of \mybrF.
There is a suggestion for $\Lambda_{\rm c}(2595)^{+}$ but not for other known $\Lcps$
in the invariant mass spectrum of $\Scpp\pim$. We see unexplained structures in the 
spectrum of $m(\Scpp\pim\pim)$ (Fig.~\ref{fig:B2Sigc:Res}).
\end{minipage}
\rule{0.02\linewidth}{0mm}
\begin{minipage}{0.48\linewidth}
\begin{center}
\vspace{0em}
\includegraphics[width=0.99\textwidth]{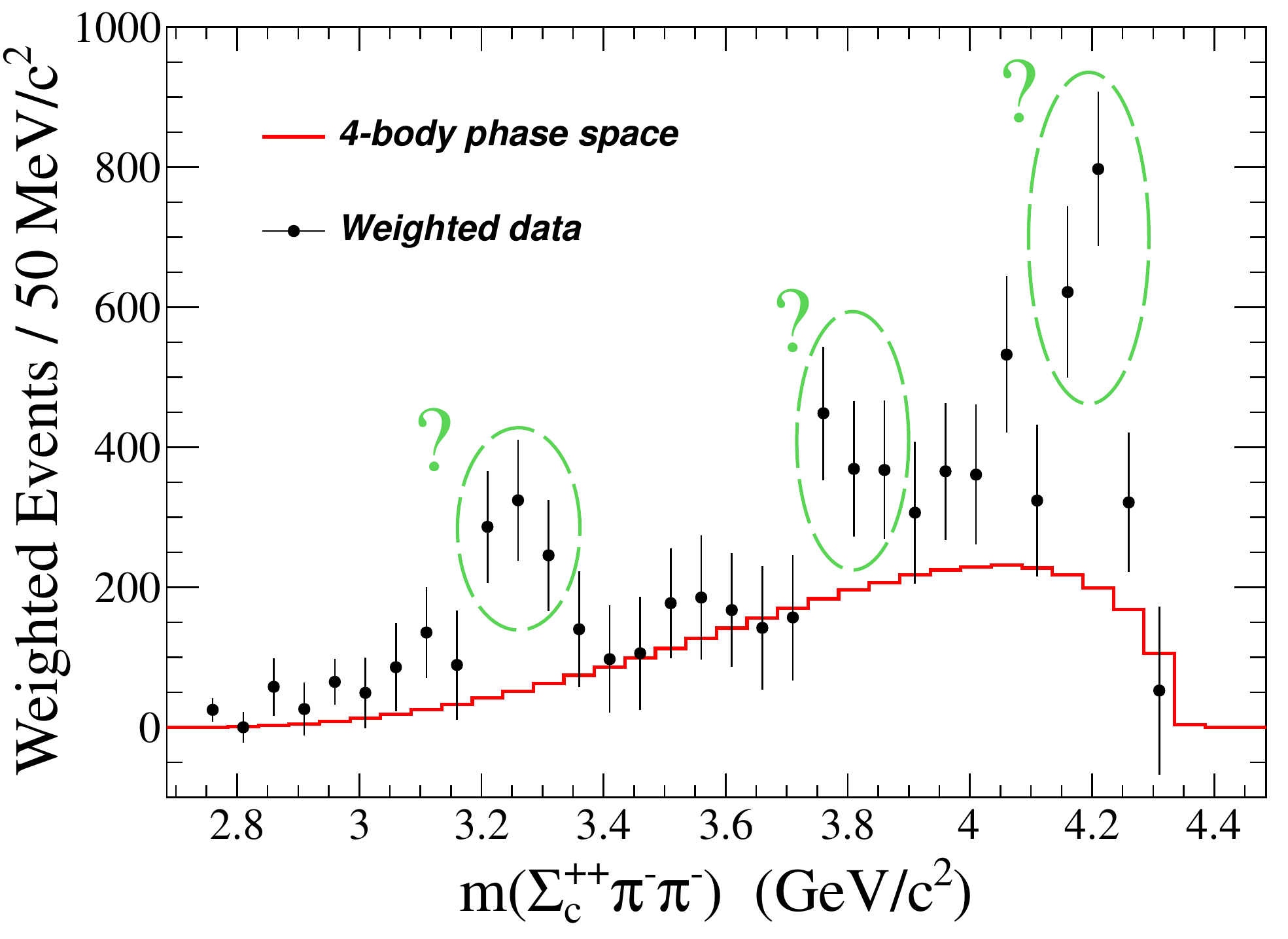}
\caption{The $m(\Scpp\pim\pim)$ distribution in data after efficiency correction and \DeltaE-sideband 
subtraction compared to simulated four-body phase space decays.}
\label{fig:B2Sigc:Res}
\end{center}
\end{minipage}
\end{figure}

\section{Threshold enhancement}

\begin{figure}[ht!]
\begin{center}
	\subfloat[]{\includegraphics[width=0.49\textwidth]{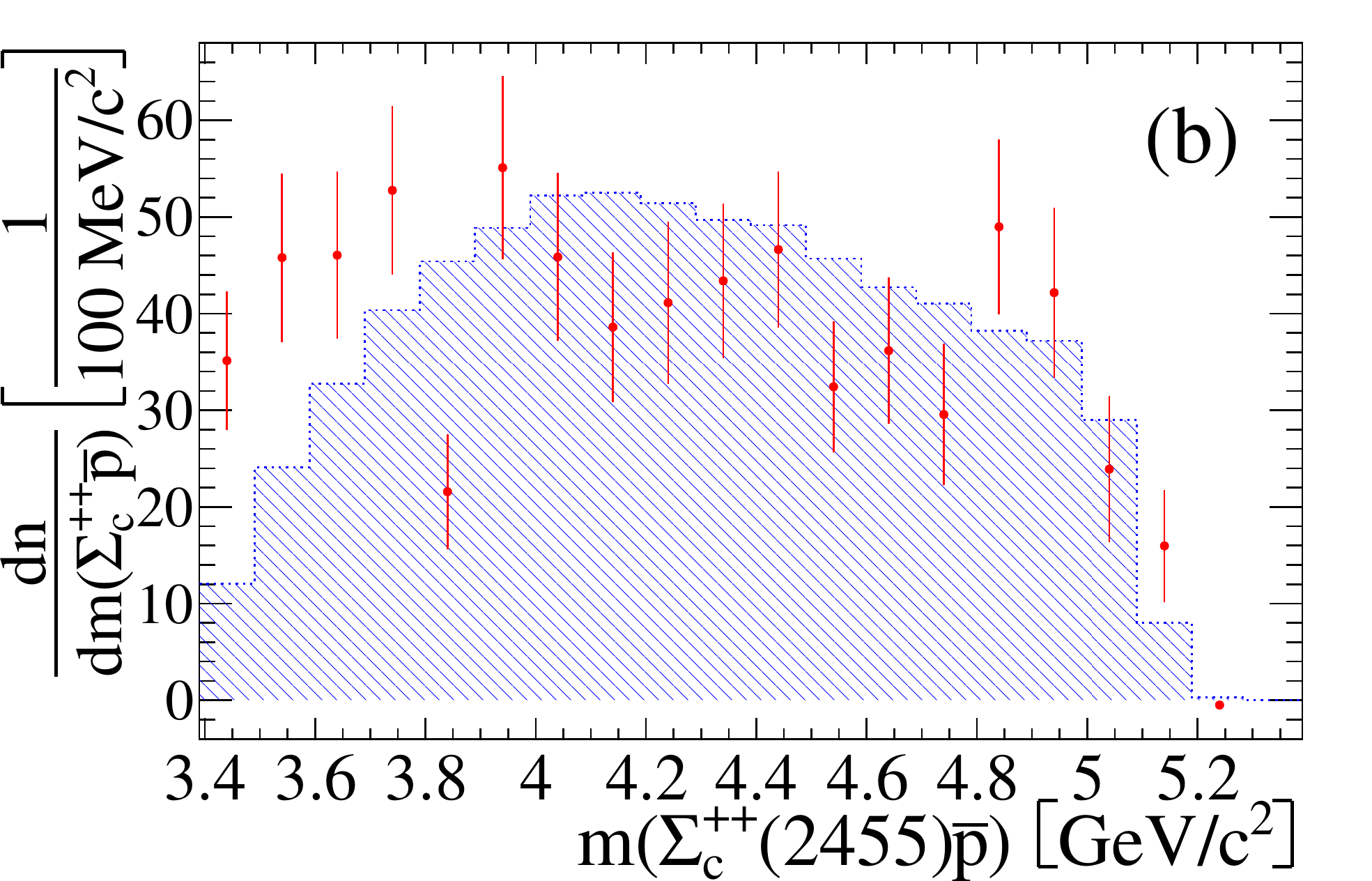}}
	\rule{2mm}{0mm}
	\subfloat[]{\includegraphics[width=0.49\textwidth]{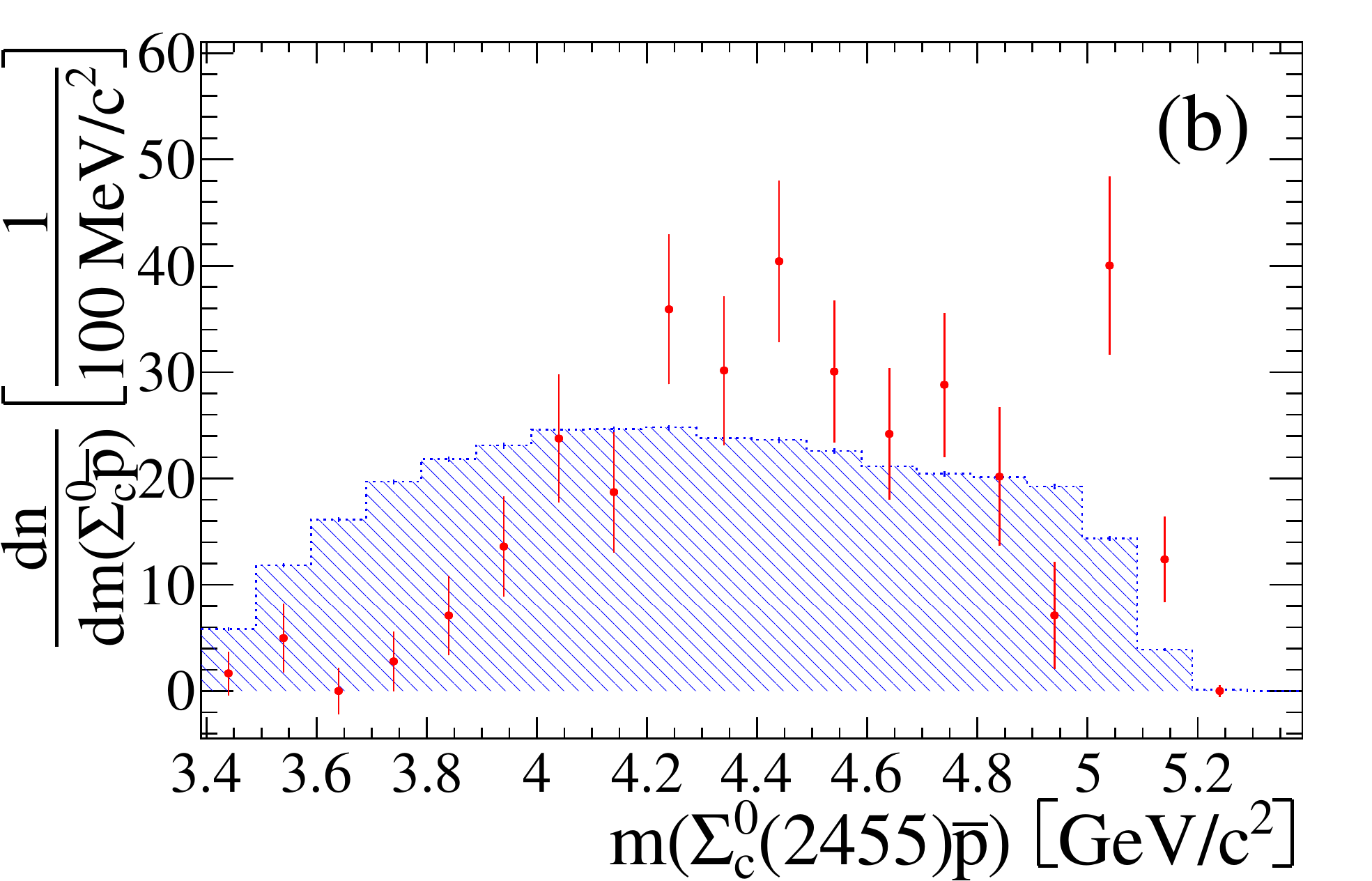}}
	\caption{The invariant baryon-antibaryon mass of the decays $\Bzb\to\Scpp(2455)\antiproton\pim$ and 
	$\Bzb\to\Scz(2455)\antiproton\pip$ for selected events from data (points with error bars) and from 
	reconstructed phase-space generated MC events of the respective decay (histogram) scaled to the same integral.}
	\vspace{-2em}
	\label{fig:B2Lc2pi:Res}
\end{center}
\end{figure}

\noindent
The threshold enhancement has been observed in a number of baryonic \B decays and may be a key 
feature in understanding hadronization processes into baryons. Figure \ref{fig:B2Lc2pi:Res} shows 
the invariant baryon-antibaryon mass of the two resonant subchannels $\Bzb\to\Scpp(2455)\antiproton\pim$ 
and $\Bzb\to\Scz(2455)\antiproton\pip$ of the decay \thdec. It can be seen that in the data sample the 
decay rate of $\Bzb\to\Scpp(2455)\antiproton\pim$ is clearly enhanced at the mass threshold 
in contrast to the distribution of $\Bzb\to\Scz(2455)\antiproton\pip$.


\clearpage

\noindent
In the past years several
theoretical models have been developed to describe this feature, e.g. baryon form factors, 
glueballs and pole models \cite{Suzuki:2006nn}. One simple ansatz to describe the threshold enhancement are QCD 
considerations according to which the partial decay rate $\Gamma_i$ is proportional to the running 
coupling constant $\as(\qsq)$ and the gluon propagator term $1/\qsq$. This leads to the effect 
that the hadronization process prefers soft gluons and produces baryon pairs with small invariant masses.

\begin{figure}[ht!]
\begin{center}
	\subfloat[]{\includegraphics[width=0.49\textwidth]{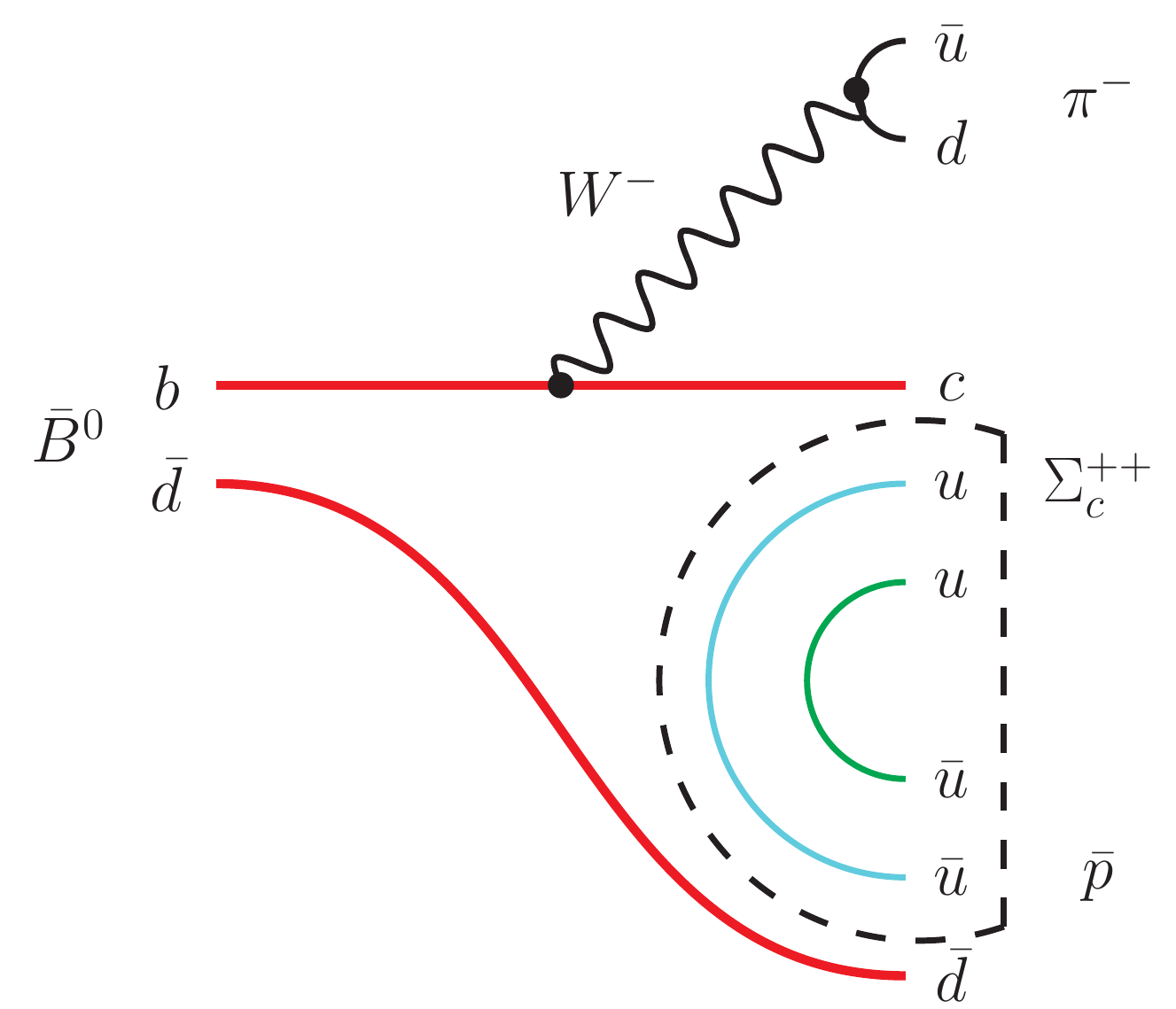}}
	\rule{2mm}{0mm}
	\subfloat[]{\includegraphics[width=0.49\textwidth]{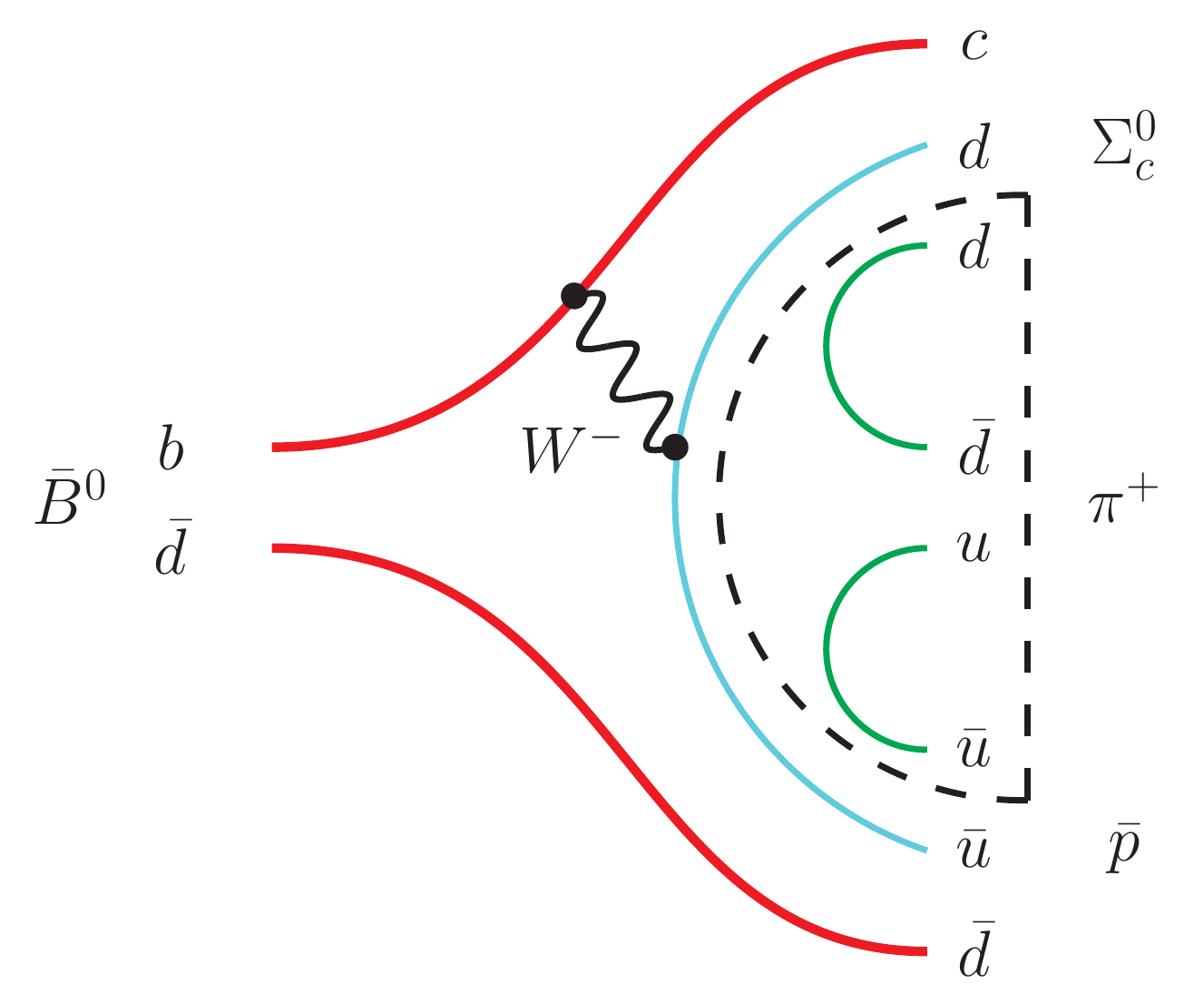}}
	\caption{One example diagram for the decay $\Bzb\to\Scpp\antiproton\pim$ and $\Bzb\to\Scz\antiproton\pip$}
	\label{fig:B2Lc2pi:FD}
\end{center}
\end{figure}

In case of the very different behaviour of the invariant baryon-antibaryon mass distribution of
$\Bzb\to\Scpp\antiproton\pim$ and $\Bzb\to\Scz\antiproton\pip$, the pole model can be used to describe 
this. Figure \ref{fig:B2Lc2pi:FD} shows one example diagram for both decays. The quarks that are enclosed
by the dashed line are created by gluons after the \Bzb decay.

\noindent
According to the pole model the left diagram represents the decay of a $\Bzb$ to a $\pim$ and a virtual 
\Dp meson pole which decays via soft gluons to a $\Scpp\antiproton$ pair with a small invariant mass. 
In contrast, the right diagram shows the decay of a $\Bzb$ to a diquark pair that hadronizes to a baryon
and a virtual baryon pole, but never to a meson and a virtual meson pole. In the end, the baryon pole decays 
to a baryon and a pion. Since the diquark content of both baryons are created back-to-back in the \Bzb rest frame, 
the $\Scz\antiproton$ pair has a high invariant mass and the extra quarks are created by hard gluons.
In combination with the QCD considerations this leads to the conclusion that the matrix element of the left diagram
is larger than the one of the right diagram resulting in a higher decay rate of $\Bzb\to\Scpp\antiproton\pim$.

In summary, the pole model and QCD considerations provide a qualitative description of the observation of an enhanced 
decay rate of $\Bzb\to\Scpp\antiproton\pim$ w.r.t. $\Bzb\to\Scz\antiproton\pip$ and the different distributions of 
the invariant baryon-antibaryon mass.

\clearpage

\section{\cvdec~\cite{christian}}

\begin{figure}[ht!]
\begin{center}
\includegraphics[width=0.6\textwidth]{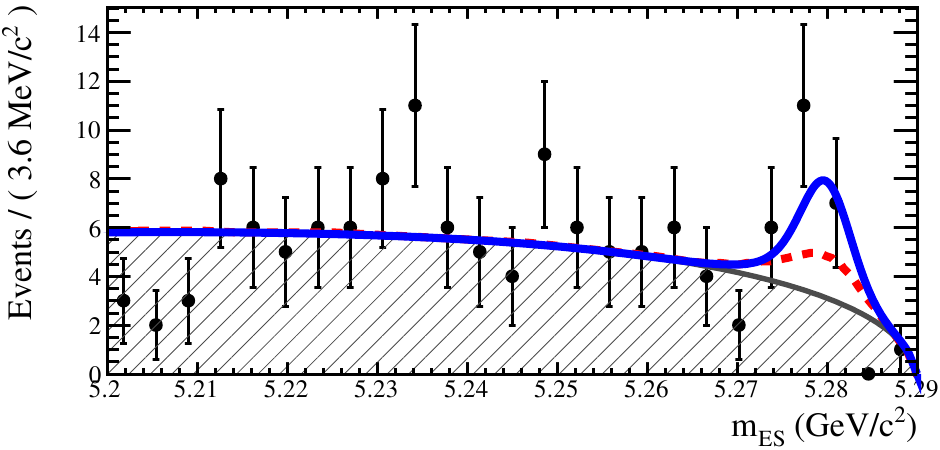}
\caption{The distribution of $m(\Dz\Lambda\Lbar)$ of selected $\cvdec$ candidates with $\Dz\to\Km\pip$
and the projection of the fit result.} 
\label{fig:B2DLL}
\end{center}
\end{figure}

\noindent
This decay can be compared to the similar decay $\Bzb\to\Dz\proton\antiproton$ and allows studies on the 
$\s\sbar$ suppression in fragmentation at low $\qsq$ in baryonic \B decays. 
From jet fragmentation results it is predicted that $N(\u\ubar):N(\s\sbar) \sim 3:1$. 
Under the assumption of an equal decay rate of the four possible final states 
$\Lambda\Lbar$, $\Lambda\Sigbar^{0}$, $\Sigma^{0}\Lbar$ and $\Sigma\Sigbar^{0}$ we expect that 
$N(\proton\antiproton):N(\Lambda\Lbar)\sim 3:\frac{1}{4} = 12 :1$.
We reconstruct $\Lambda\to\proton\pim$ and \Dz in the final states $\Km\pip, \Km\pip\piz$ 
and $\Km\pip\pip\pim$. The number of signal events is determined simultaneously in an unbinned two-dimensional fit to $m(\Dz\Lambda\Lbar)$ and $\mes$ 
of all \Dz subsamples. In addition, contributions from $\Sigma\to\Lambda\g$ decays are accounted for in the fit model. 
Figure \ref{fig:B2DLL} shows the distribution of $m(\Dz\Lambda\Lbar)$ of the $\Bzb$ candidates from the $\Dz\to\Km\pip$ subsample and 
the fit result.

\noindent
The branching fraction is determined to be $\cvbr = (9.8^{+2.9}_{-2.6}\,\ustat\pm1.9\usyst)\times10^{-6}$ corresponding to a 
significance of $3.4\sigma$. In combination with the result $\BR(\Bzb\to\Dz\proton\antiproton)=(1.04\pm0.07)\times10^{-4}$ \cite{RPP}
this leads to $ \frac{\BR(\Bzb\to\Dz\proton\antiproton)}{\cvbr} = 10.6 \pm 3.8 $, which is compatible with the aforementioned 
prediction of $12:1$.

%

\end{document}